\documentclass[12pt]{iopart}
\usepackage{iopams}
\eqnobysec
\usepackage{graphicx}	
\expandafter\let\csname equation*\endcsname\relax
\expandafter\let\csname endequation*\endcsname\relax
\usepackage{amsmath}	
\usepackage{iopams}
\usepackage{notoccite}
\usepackage{aas_macros}
\usepackage{soul}
\usepackage{xcolor}

\begin{document}

\title[Radial stability of ultra compact Schwarzschild stars]{On the radial stability of ultra compact Schwarzschild stars beyond the Buchdahl limit}

\author{Camilo Posada$^1$ and Cecilia Chirenti$^2$}

\address{$^{1}$ \emph{Institute of Physics and Research Centre of Theoretical Physics and Astrophysics, Faculty of Philosophy and Science, Silesian University in Opava, Bezru\v{c}ovo n\'{a}m. 13, CZ-74601 Opava, Czech Republic}}
\address{$^{2}$ \emph{Center for Mathematics, Computation and Cognition, UFABC, Av. dos Estados, 5001, Santo Andr\'e - SP, 09210-580, Brazil}}
\vspace{10pt}
\date{}

\begin{abstract}

In this paper we used the theory of adiabatic radial oscillations developed by Chandrasekhar to study the conditions for dynamical stability of constant energy-density stars, or Schwarzschild stars, in the unstudied ultra compact regime beyond the Buchdahl limit, that is, for configurations with radius $R$ in the range $R_{\rm S}<R<(9/8)R_{\rm S}$, where $R_{\rm S}$ is the Schwarzschild radius of the star. These recently found analytical solutions exhibit a negative pressure region in their centre and, in the limit when $R\to R_{\rm S}$, the full interior region of the star becomes filled with negative pressure. Here we present a systematic analysis of the stability of these configurations against radial perturbations. We found that, contrary to the usual expectation found in many classical works, the ultra compact Schwarzschild star is stable against radial oscillations. We computed values of the critical adiabatic index $\gamma_{c}$ for several stellar models with varying radius $R/R_{\rm S}$ and found that it also approaches a finite value as $R/R_{\rm S} \to 1$.

\end{abstract}

%
\vspace{2pc}
\noindent{\it Keywords\/}: Ultra compact stars, radial oscillations, stellar stability, interior solution
%
%
%
%
\section{Introduction}

In two seminal works published in 1964, Chandrasekhar \cite{chandra1964a,chandra1964b} developed the theory of radial oscillations of relativistic stars, which has had numerous applications to the characterization of stable and unstable branches of different compact stellar models. The dynamic equation governing radial pulsations leads to a self-adjoint eigenvalue problem, which can be solved using several methods \cite{bardeen1966}. Assuming that the adiabatic index $\gamma$ of the perturbation is constant throughout the star, Chandrasekhar used trial eigenfunctions to determine the critical values $\gamma = \gamma_{\rm c}$ for the onset of instability, for stars described by the Schwarzschild constant-density interior solution, or Schwarzschild stars (see, e.g.~\cite{wald1984,stuchlik2000}) and for relativistic polytropes. One result of particular interest is that for a Schwarzschild star with total mass $M$ and radius $R$, in the Newtonian limit of low compactness $M/R$, we have $\gamma_{\rm c} = 4/3 + (19/42)R/R_{\rm S}$, where $R_{\rm S} = 2M$ is the Schwarzschild radius of the star.

It is well known that the central pressure of the Schwarzschild star diverges for configurations obtained at the end of a sequence with increasing compactness when its radius reaches $R=(9/4)M$. The relevance of this limit relies on the fact that its existence is not restricted to the case of homogeneous density stars. As it was shown by Buchdahl \cite{buchdahl1959}, if one assumes that the energy density $\epsilon$ is positive everywhere inside the star and decreases monotonically with $r$, namely $d\epsilon/dr\leq0$, then it is possible to find a general upper mass limit, $M\leq 4R/9$, which will be independent of the equation of state. Thus the Schwarzschild star provides an extreme case which saturates the Buchdahl limit; as a result, configurations more compact than the Buchdahl bound have been usually dismissed as being unstable.

Nevertheless, when one considers the Schwarzschild star beyond the Buchdahl limit, namely $R_{S}<R<(9/8)R_{\rm S}$, the solutions present an interesting behavior \cite{mazur2015}. The pole where the pressure diverges moves out from the centre up to a surface of radius $R_{0}=3R\sqrt{1-\frac{8}{9}\frac{R}{R_{\rm S}}} < R$, and a regular negative pressure interior region emerges in the range $0<r<R_{0}$ while the pressure is still positive in the range $R_0 < r < R$ \footnote{The interior negative pressure region is matched to the outer portion of the star (or to the exterior vacuum in the limiting case) through an infinitesimal surface-layer of transverse stresses endowed with a surface tension proportional to the difference in the surface gravities.}. In the ultra compact limit when $R_{0}\to R_{\rm S}^{-}$ from inside and $R\to R_{\rm S}^{+}$ from outside, the whole interior of the star has negative pressure determined by a modified de Sitter space-time.  Thus, in this limiting case the ultra compact Schwarzschild star becomes essentially the gravastar proposed by Mazur and Mottola \cite{mazur2001,mazur2004}.

The gravastar model has attracted considerable interest as an alternative to black holes since it provides an ultra compact non-singular object with no event horizon. Variations of the model inspired many related solutions, for instance the continuous pressure models described in \cite{cattoen2005,benedictis2006} and the solutions supported by nonlinear electrodynamics given by \cite{lobo2006}. Observational signatures of the model were studied as early as 2007 with an analysis of its gravitational quasinormal mode spectrum \cite{chirenti2007}, and more recently in \cite{volkel2017}. Late works like \cite{sakai2014,pani2015} have considered generalizations of black hole shadows and the I-Love-Q relations \cite{yagi2013} in this context.

The stability of many different proposed black hole alternatives inspired by the original gravastar model has been widely studied in the literature, as stability is essential for any configuration to be considered as a physically viable model. Stability against radial oscillations was analyzed for an infinitesimally thin shell model \cite{visser2003} and configurations with continuous pressures \cite{horvat2011}. The existence of stable configurations makes them possible candidates for ``black hole mimickers".

After the first recent gravitational-wave detections announced by LIGO \cite{abbott2016a} we finally have the possibility to probe gravity closer to the horizon scale \cite{abbott2016b} and start constraining alternative models \cite{chirenti2016}. Thus in the future, we expect to see possible deviations from general relativity due to quantum corrections  \cite{cardoso2017}. In principle, these quantum corrections could be responsible for the formation of gravastar-like objects instead of classical black holes. However, more stringent tests of this scenario may have to wait for observations with space detector LISA \cite{gair2013} or third generation ground-based gravitational wave detectors such as the Einstein Telescope \cite{punturo2010} or the Cosmic Explorer \cite{abbott2017}.

Therefore, motivated by the aforementioned works, in this paper we present a systematic analysis of the conditions for dynamical instability of constant-density stars, or Schwarzschild stars, in the unexplored regime $R_{S}<R<(9/8)R_{\rm S}$. Our results can be considered as an extension of those presented by Chandrasekhar \cite{chandra1964a,chandra1964b}, and we show that, quite surprisingly, stability is regained after the classical instability at $R = (9/8)R_{S}$, in the sense that we find that stable configurations are allowed below the Buchdahl bound. In particular, we show that configurations with $R \to R_{S}$ are also stable.

This paper is organized as follows: in section \ref{sect2} we review briefly the Schwarzschild interior solution, or Schwarzschild star, in the negative pressure regime for higher compactness approaching the Schwarzschild black hole limit. In section \ref{sect3} we outline the pulsation equation for radial oscillations derived by Chandrasekhar in the framework of general relativity. We specialize the problem to the case of a constant energy-density configuration in section \ref{sect4}. Finally, in section \ref{sect5} we present our methods and results and in section \ref{sect6} we provide the conclusions. Throughout the paper, we use geometrized units where $G=c=1$. 

\section{The Schwarzschild star and negative pressure interior}\label{sect2}

In this section we briefly summarize the Schwarzschild interior solution, or Schwarzschild star, corresponding to a uniform-density spherical star, in the regime of high compactness with a negative pressure interior (for a more detailed mathematical description of this solution, see \cite{mazur2015,posada2017}). The starting point is a spherically symmetric spacetime in the standard Schwarzschild form 

\begin{eqnarray}\label{metric0}
ds^2 = -e^{2\nu}dt^2 + e^{2\lambda}dr^2 + r^2\left(d\theta^2 + \sin^2\theta d\phi^2\right).
\end{eqnarray} 

\noindent We model the star as a spherically symmetric fluid configuration characterized by a stress-energy tensor with the form

\begin{eqnarray}\label{fluid}
T_{\;\;\nu}^{\mu} = 
\begin{pmatrix}
-\epsilon & 0 & 0 & 0\\
0 & p & 0 & 0\\
0 & 0 & p_{\perp} & 0\\
0 & 0 & 0 & p_{\perp}
\end{pmatrix} 
\end{eqnarray}

\noindent where $\epsilon$ is the total mass-energy density, $p$ is the radial pressure and $p_{\perp}$ corresponds to the transversal pressure. The relation between the energy density $\epsilon$ and the pressure $p$ will be determined once an equation of state of the form $p=p(\epsilon)$ is given. The Einstein equations together with the energy-momentum conservation relation provide the pair of equations which determine the relativistic hydrostatic equilibrium condition

\begin{eqnarray}\label{tov1} 
\frac{dp}{dr} + (\epsilon+p)\frac{d\nu}{dr}+\frac{2}{r}(p-p_{\perp})=0,
\end{eqnarray}

\noindent and

\begin{eqnarray}\label{tov2}
m(r) \equiv 4\pi\int_{0}^{r}\epsilon~r^2dr.
\end{eqnarray}

\noindent At the center of the star, $p(r=0)=p_{c}=\text{constant}$ and $m(0)=0$. At the surface of the star, $p(R)=0$, and $m(R)=M$ is the total  mass of the star.  The total mass-energy density $\epsilon$ is constant everywhere inside the star and, according to \eqref{tov2}, it is given by

\begin{eqnarray}
\epsilon = \frac{3M}{4\pi R^3}.
\end{eqnarray}

\noindent Thus, \eqref{tov1} and \eqref{tov2} can be integrated analytically. We introduce the following parametrization \cite{chandra1964b,chandra1974} 

\begin{eqnarray} \label{param}
y^2 = 1-\left(\frac{r}{\alpha}\right)^2\quad {\rm and}\quad y_{1}^2 = 1-\left(\frac{R}{\alpha}\right)^2,
\end{eqnarray}  

\noindent where $r$ is measured in units of $\alpha$, which is defined as

\begin{eqnarray} \label{alpha}
\alpha^2\equiv\frac{3}{8\pi\epsilon}=\frac{R^3}{R_{\rm S}}.
\end{eqnarray}  

\noindent In terms of the parametrization \eqref{param}, the Schwarzschild interior solution takes the form

\begin{eqnarray}\label{pressure}
p=\epsilon\left(\frac{y-y_{1}}{3y_{1}-y}\right),
\end{eqnarray}  
\begin{eqnarray}\label{interior}
e^{2\lambda}=\frac{1}{y^2}\quad {\rm and} \quad e^{2\nu}=\frac{1}{4}(3y_{1}-y)^2.
\end{eqnarray}

\noindent The central pressure can be obtained from \eqref{pressure}, and it is given by

\begin{eqnarray}\label{centralp}
p_{\rm c}=\epsilon\left(\frac{1-y_{1}}{3y_{1}-1}\right),
\end{eqnarray}  

\noindent which is a constant. The Buchdahl condition $R/R_{\rm S}>9/8$, equivalent to $3y_{1}>1$ in this parametrization, guarantees that the pressure is positive and finite everywhere inside the star \cite{buchdahl1959}. At the stellar surface $r=R$, the interior solution must match the exterior Schwarzschild solution $e^{2\nu}=e^{-2\lambda}=1-2M/r$. Note that the pressure \eqref{pressure} is regular everywhere inside the star, except at some radius $R_{0}$ where the denominator vanishes, which is a first-order pole given by

\begin{eqnarray}\label{pole}
R_{0}=3R\sqrt{1-\frac{8}{9}\frac{R}{R_{\rm S}}}<R,
\end{eqnarray}

\noindent which is imaginary for $R/R_{\rm S}>9/8$. Moreover,  as \eqref{pressure} and \eqref{interior} show, the pressure diverges exactly at the same point where $e^{2\nu}=0$.

\begin{figure}
\centering
\includegraphics[scale=0.6]{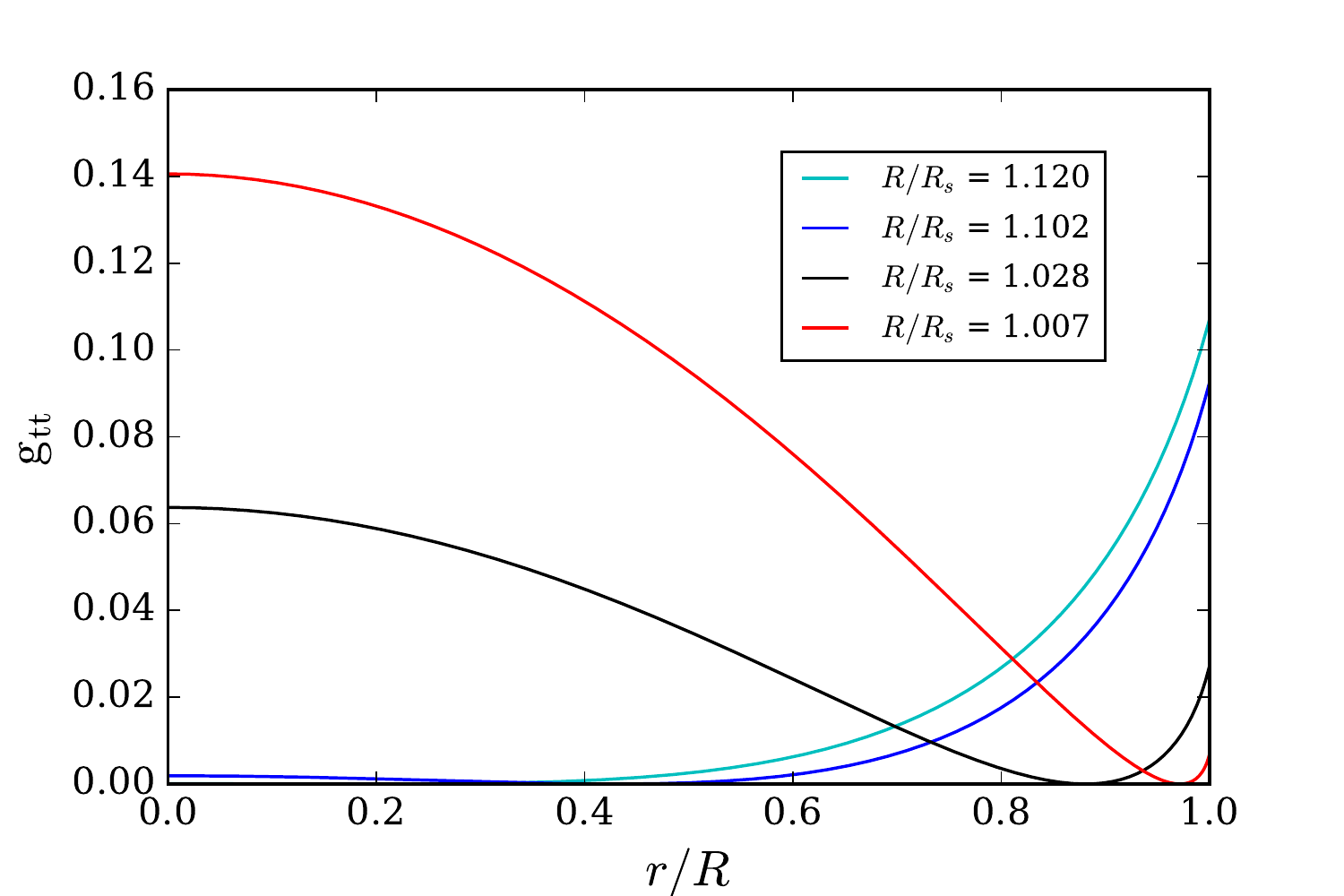}
\caption{\label{fig1} Metric component $\mathrm{g_{tt}=e^{2\nu}}$ as a function of $r$ (in units of the radius of the star $R$) of the Schwarzschild star for different values of the parameter $R/R_{S}<9/8$.}
\end{figure}

\begin{figure}
\centering
\includegraphics[scale=0.6]{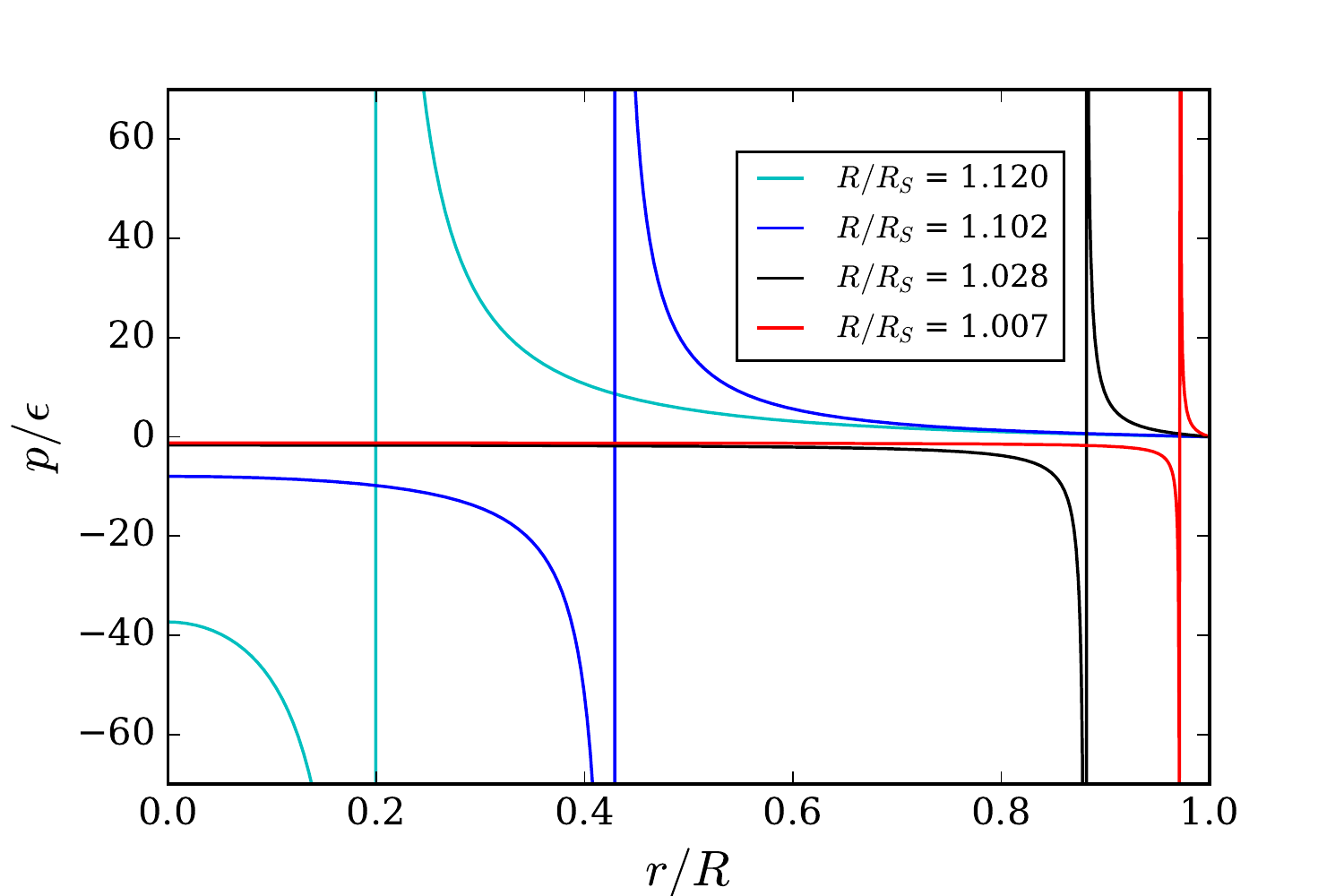}
\caption{\label{fig2} Pressure (in units of $\epsilon$) as a function of $r$ (in units of the radius of the star $R$) of the Schwarzschild star for the same values of the parameter $R/R_{S}$ as in \fref{fig1}. Note how the ratio $p/\epsilon\to -1$, as the radius of the star $R$ approaches the black hole limit.}
\end{figure}

Despite this seemingly unphysical behavior for the pressure inside the star, in the regime $1<R/R_{\rm S}<9/8$, a further analysis allows us to uncover some interesting features. For increasingly compact stars in this regime, the pole in the pressure \eqref{pole} moves from the center of the star to finite values $0<R_{0}<R$ (see figure 1 in \cite{posada2017}). This creates a new region with negative pressure in the interval $0\leq r < R_{0}$, but $e^{2\nu}>0$, thus $g_{tt}$ remains timelike (see \fref{fig1} and \fref{fig2}). Thus, in principle, if we could represent a collapsing star by a quasistatic series of increasingly more compact equilibrium configurations, we would see it undergo a phase transition which occurs first at the center and then spreads out towards the surface.

As the radius of the star approaches the Schwarzschild radius from above, $R\to R_{\rm S}^{+}$, the radius of the negative pressure region approaches it from below, $R_{0}\to R_{\rm S}^{-}$. In the ultra compact limit when $R=R_{0}=R_{\rm S}$, the Schwarzschild interior solution \eqref{pressure} and \eqref{interior} shows that the whole new interior region is described by the equation of state $p=-\epsilon$ for $r<R=R_{0}=R_{\rm S}$. In this ultra compact limit, the interior metric functions given in \eref{interior} describe a patch of a modified de Sitter spacetime

\begin{eqnarray}
e^{2\nu}=\frac{1}{4}(1-H^2r^2)=\frac{1}{4}e^{-2\lambda}\quad {\rm with} \quad H=\frac{1}{R_{\rm S}}.
\end{eqnarray}  

\noindent The interior is perfectly regular at $r=0$, in contrast to the singular behavior of black holes at that point. Outside the star (for $r>R_{\rm S}$), the space-time geometry is determined by the spherically symmetric exterior Schwarzschild metric. The Schwarzschild surface at $r=R_{\rm S}$ is a marginally trapped surface, but there is no interior trapped surface for $r<R_{\rm S}$, and no event horizon.

Finally, there is an infinitesimal thin-shell discontinuity at $r=R_{0}=R_{\rm S}=2M$ where the interior and the exterior metrics must be matched by using the Israel junction conditions \cite{israel1966}. The divergence in the pressure at the surface $R_{0}$ can be regularized through the Komar integral \cite{komar1959} together with the relaxation of the isotropic fluid condition at $r=R_{0}$ (see \cite{mazur2015} for details). The anisotropy in the pressure at the surface $R_{0}$ produces a surface tension which is proportional to the difference in surface gravities. Thus, in the ultra compact limit when $R = R_{0} = R_{\rm S}$, the Schwarzschild star becomes essentially the non-singular gravitational condensate star or gravastar proposed in \cite{mazur2001,mazur2004}.

\section{Equations for infinitesimal radial oscillations}
\label{sect3}

The equation that governs radial oscillations in the fluid of a relativistic star was derived by Chandrasekhar \cite{chandra1964b} using variational methods, under the following assumptions: the spherical symmetry \eqref{metric0} of the configuration is not affected by the perturbations, there are no dissipative effects and the baryon number is conserved. Following Chandrasekhar, the dynamical stability of the star can be studied through the analysis of the normal modes of radial oscillations. Assuming that the perturbations have the time-dependent form $e^{-i\omega t}$, the normal modes will be determined by the pulsation equation

\begin{eqnarray} \label{pulsation}
\fl \omega^2e^{2(\lambda-\nu)}(\epsilon+p)\xi = \frac{4}{r}\left(\frac{dp}{dr}\right)\xi-e^{-(\lambda+3\nu)}\frac{d}{dr}\left[e^{(\lambda+3\nu)}\left(\frac{\gamma~p}{r^2}\right)\frac{d}{dr}(r^2e^{-\nu}\xi)\right] \nonumber\\
 +8\pi~e^{2\lambda}p(\epsilon+p)\xi-\frac{1}{\epsilon+p}\left(\frac{dp}{dr}\right)^2\xi,
\end{eqnarray}

\noindent where $\xi$ corresponds to the Lagrangian displacement, or in this case the radial displacement of a fluid element from its equilibrium position, $\omega$ denotes the characteristic frequency of the oscillation, and

\begin{eqnarray} \label{gamma}
  \gamma=\frac{\epsilon+p}{p}\left(\frac{dp}{d\epsilon}\right)_{S}
\end{eqnarray}

\noindent is the adiabatic index associated to the perturbations, or `effective' $\gamma$ \cite{merafina1989}. The subscript S in \eqref{gamma} indicates at constant specific entropy. Note that this $\gamma$ is not necessarily equal to the adiabatic index, or `ratio of specific heats' associated to the equation of state of the star. For instance, for the Schwarzschild star with constant-density, the adiabatic index $\gamma$ for the fluid is infinite (incompressible fluid). However, for the perturbations, constant finite values of the `effective' $\gamma$ are being assumed \cite{chandra1964b,shapiro2007}.

Physically acceptable solutions of \eqref{pulsation} must satisfy the following boundary conditions

\begin{eqnarray} \label{bc1}
\eqalign
\xi = 0\quad \text{at}\quad r=0 \label{bc1a},\\
\Delta p = 0\quad \text{at}\quad r=R. \label{bc1b}
\end{eqnarray}

\noindent Note that \eqref{bc1a} guarantees that there is no fluid motion in the center, and \eqref{bc1b} indicates that the pressure vanishes at the surface of the star. The stability condition is established through the variational method \cite{chandra1964b}. Following this principle, the condition for the onset of dynamical instability of the star is that the RHS of 

\begin{eqnarray} \label{eqradial2}
\omega^2\int_{0}^{R}dr~e^{(3\lambda-\nu)/2}(\epsilon+p)r^2\xi^2 = 4\int_{0}^{R}dr~e^{(\lambda+\nu)/2}r\left(\frac{dp}{dr}\right)\xi^2 \nonumber\\
+\int_{0}^{R}dr~e^{(\lambda+3\nu)/2}\frac{\gamma~p}{r^2}\left[\frac{d}{dr}(r^2e^{-\nu/2}\xi)\right]^2 - \int_{0}^{R}dr~e^{(\lambda+\nu)/2}\left(\frac{dp}{dr}\right)^2\frac{r^2\xi^2}{\epsilon+p} \nonumber\\
+8\pi\int_{0}^{R}dr~e^{(3\lambda+\nu)/2}p(\epsilon+p)r^2\xi^2,
\end{eqnarray}

\noindent vanishes for some function $\xi$, given the boundary conditions \eqref{bc1a} and \eqref{bc1b}. For imaginary values of $\omega$, i.e. $\omega^2<0$, the perturbation $e^{-i\omega t}$ grows exponentially in time, and the configuration will be unstable. Solutions purely oscillatory, i.e., $\omega^2>0$, will be stable. Thus $\omega^2=0$, also known as the \emph{neutral mode}, separates stable and unstable models  \cite{shapiro2007}. Following \cite{mtw} we introduce the renormalized displacement function

\begin{eqnarray} \label{zeta}
\zeta\equiv r^2e^{-\nu}\xi.
\end{eqnarray}

\noindent In terms of $\zeta$, \eqref{pulsation} can be written in the self-adjoint form

\begin{eqnarray}\label{radialeq}
\frac{d}{dr}\left(P\frac{d\zeta}{dr}\right)+\left(Q+\omega^2W\right)\zeta=0,
\end{eqnarray}

\noindent where

\begin{eqnarray}\label{p}
P \equiv \gamma~r^{-2}p~e^{\lambda+3\nu},
\end{eqnarray}
\begin{eqnarray}\label{q}
Q \equiv r^{-2}~e^{\lambda+3\nu}(\epsilon+p)\left[\left(\frac{d\nu}{dr}\right)^2+\frac{4}{r}\left(\frac{d\nu}{dr}\right)-8\pi~e^{2\lambda}p\right],
\end{eqnarray}
\begin{eqnarray}\label{w}
W \equiv (\epsilon+p)r^{-2}~e^{3\lambda+\nu},
\end{eqnarray}

\noindent and the boundary conditions \eqref{bc1a} and \eqref{bc1b} imply that

\begin{eqnarray} \label{bc2}
\eqalign
\zeta(0) = 0, \label{bc2a}\\
\gamma p\left(\frac{d\zeta}{dr}\right)\Bigg|_{R} = 0. \label{bc2b}
\end{eqnarray}

\noindent Thus \eqref{radialeq}, together with the boundary conditions \eqref{bc2a} and \eqref{bc2b}, correspond to a self-adjoint eigenvalue problem for the frequencies $\omega^2$. Given the mathematical features of this problem, the following results apply \cite{shapiro2007}: the frequencies $\omega^2$ are real and form a discrete spectrum $\omega_{0}^2<\omega_{1}^2<\omega_{2}^2<\cdot\cdot\cdot<\omega_{n}^2$. Therefore if the fundamental mode is stable, i.e., $\omega_{0}^2>0$, then all radial modes are stable. The eigenfunction $\xi_{n}$, which corresponds to the $n$-th mode $\omega_{n}^2$, has $n$ nodes in the interval $0<r<R$. Furthermore, the $\xi_{n}$ are orthogonal.      

The methods to solve the eigenvalue problem \eqref{radialeq} are well known and have been discussed in \cite{bardeen1966}. For instance, by using trial functions, Chandrasekhar \cite{chandra1964b} found the critical value $\gamma_{\rm c}$ of the adiabatic index for the dynamical stability of the constant energy-density star, or Schwarzschild star, for different values of the compactness. Thus, the `effective' $\gamma$ must be greater than $\gamma_{c}$ for the configuration to be stable against radial perturbations. When relativistic effects are small, Chandrasekhar found that   

\begin{eqnarray} \label{gammach}
\gamma_{\rm c} = \frac{4}{3} + \frac{19}{42}\frac{R_{\rm S}}{R}, \quad {\rm for} \quad M/R \ll 1.
\end{eqnarray}

\noindent If $\gamma<\gamma_{\rm c}$, dynamical instability will ensue. Thus, \eqref{gammach} corresponds to the relativistic correction of the Newtonian value $\gamma_{\rm N}=4/3$. In the next sections we will study the radial stability of an ultra compact Schwarzschild star in the negative pressure regime $R_{\rm S}<R<(9/8)R_{\rm S}$.

\section{Radial stability of ultra compact Schwarzschild stars}
\label{sect4}

Our particular interest is to study the radial stability of the ultra compact Schwarzschild star. It is convenient to introduce the new independent variable

\begin{eqnarray} \label{x}
x \equiv 1 - y = 1-\left[1-\left(\frac{r}{\alpha}\right)^2\right]^{1/2},
\end{eqnarray}

\noindent such that $x\in(0,1-y_{1})$, with $y_{1}^2=1-(R/\alpha)^2$. In terms of the new variable $x$, \eqref{radialeq} adopts the form

\begin{eqnarray} \label{radialx}
\frac{d}{dx}\left[\frac{\sqrt{x(2-x)}}{(1-x)}\left(P\frac{d\zeta}{dx}\right)\right]+\alpha^2\frac{(1-x)}{\sqrt{x(2-x)}}(Q+\omega^2W)\zeta=0,
\end{eqnarray}

\noindent where

\begin{eqnarray}\label{p}
P(x)=\frac{\gamma\epsilon}{24\alpha^2}\frac{(k+x)^2}{x(1-x)(2-x)}\left[3(1-x)-(k+1)\right],
\end{eqnarray}
\begin{eqnarray}\label{q}
  \fl & Q(x) = \frac{\epsilon}{3\alpha^4}\frac{(k+1)(k+x)}{x(1-x)^2(2-x)}\times \nonumber\\
  & \left[1+\frac{x(2-x)}{4(1-x)(k+x)}-\frac{2\pi\epsilon\alpha^2}{3}\frac{3(1-x)-(k+1)}{(1-x)}\right],
\end{eqnarray}
\begin{eqnarray}\label{w}
W(x)=\frac{\epsilon}{3\alpha^2}\frac{(k+1)}{x(1-x)^3(2-x)},
\end{eqnarray}

\noindent and $k \equiv \vert 3y_{1}-1\vert$, where the absolute value guarantees that $\sqrt{g_{tt}}=e^{\nu}$, given by \eref{interior}, remains positive when the radius of the star goes beyond the Buchdahl bound (see the discussion in \cite{posada2017}). As usual when solving numerically second order differential equations, we first turn \eref{radialx} into two coupled first-order differential equations. Introducing

\begin{eqnarray}\label{eta}
\eta(x)\equiv\frac{\sqrt{x(2-x)}}{(1-x)}\left[P(x)\frac{d\zeta}{dx}\right],
\end{eqnarray}

\noindent we obtain

\begin{eqnarray}\label{dif1}
\frac{d\zeta}{dx}=\frac{24\alpha^2}{\gamma\epsilon}\frac{(1-x)^2\sqrt{x(2-x)}}{(k+x)^2\left[3(1-x)-(k+1)\right]}\eta(x),
\end{eqnarray}
\begin{eqnarray}\label{dif2}
\frac{d\eta}{dx}= -\alpha^2\frac{(1-x)}{\sqrt{x(2-x)}}\left[Q(x)+\omega^2W(x)\right]\zeta(x).
\end{eqnarray}

\noindent Now the boundary conditions \eqref{bc2a}--\eqref{bc2b} read

\begin{eqnarray} 
\eqalign
\zeta(x=0) = 0, \label{bc3a}\\
\eta(x=x_1) = 0, \label{bc3b}
\end{eqnarray}

\noindent where $x_1 \equiv 1-y_1$. In the next section we will discuss our numerical methods for solving the eigenvalue problem described by \eqref{dif1}--\eref{dif2}, with the boundary conditions \eqref{bc3a}--\eqref{bc3b}, and present our results.

\section{Numerical results}
\label{sect5}

\subsection{Methods}

We find the critical adiabatic index $\gamma_{\rm c}$, for the onset of instability, by setting $\omega^2=0$ in \eqref{dif2}. The resulting eigenvalue problem for $\gamma$ is solved with a \emph{shooting method}. We integrate the system \eqref{dif1}--\eqref{dif2} from the center up to the surface of the star with some trial value of $\gamma$, and iterate until we find the optimal value of $\gamma$ such that the solution satisfies (within a prescribed error) the boundary conditions \eqref{bc3a}--\eqref{bc3b}. Thus, the optimal $\gamma$ corresponds to the critical adiabatic index. We used two different methods for finding $\gamma_{\rm c}$: the binary search method, or bisection method, and the secant method \cite{press1992,newman2013}. The results of $\gamma_{\rm c}$ obtained with both methods were in agreement up to 6 decimal places.      

The equations derived in section~\ref{sect4} were integrated for different values of the parameter $R/R_{\rm S}$. The integrations were performed using a Runge-Kutta-Fehlberg (RKF) method with an adaptive stepsize \cite{press1992}. We found that the adaptive RKF method provided fast and stable results in the regime near and below the Buchdahl radius, where the equation of state is stiff. Even though the central pressure, as given by \eqref{centralp}, has a divergence at $r=0$ when $y_{1}=1/3$, or $k=0$, note that \eqref{dif1}--\eqref{dif2} are well behaved there. Thus, the only poles are located at the ends $\{0,1-y_{1}\}$, which are excluded from the integration interval, i.e., $x\in(0,1-y_{1})$.

\subsection{Radial stability above the Buchdahl bound}

As an initial code check, we computed values of $\gamma_{\rm c}$ for a Schwarzschild star in the regime $R/R_{\rm S}>9/8$. Our results are shown in \fref{fig3} (see also \tref{table1} in the Appendix), where we plot $\gamma_{\rm c}$ as a function of the ratio $R/R_{\rm S}$. We find that our results are in very good agreement with those reported by Chandrasekhar \cite{chandra1964b} and Wright \cite{wright1964}, also presented in the figure for comparison. We note that  $\gamma_{\rm c}$ approaches the Newtonian value $4/3$ when $R/R_{\rm S}\to\infty$, as expected (in the Newtonian theory, the value of $\gamma_{\rm c}$ is independent of the radius of the star). For values of $R/R_{\rm S}$ near the Buchdahl bound, $\gamma_{\rm c}$ grows rapidly indicating the approach to the instability point. Nevertheless, as we will show in section \ref{sect5.3}, this changes when one considers the negative pressure regime, $R_{\rm S}<R<(9/8)R_{\rm S}$, where the critical $\gamma_{\rm c}$ values decrease monotonically and approach a finite value.

\begin{figure}
\centering
\includegraphics[scale=0.6]{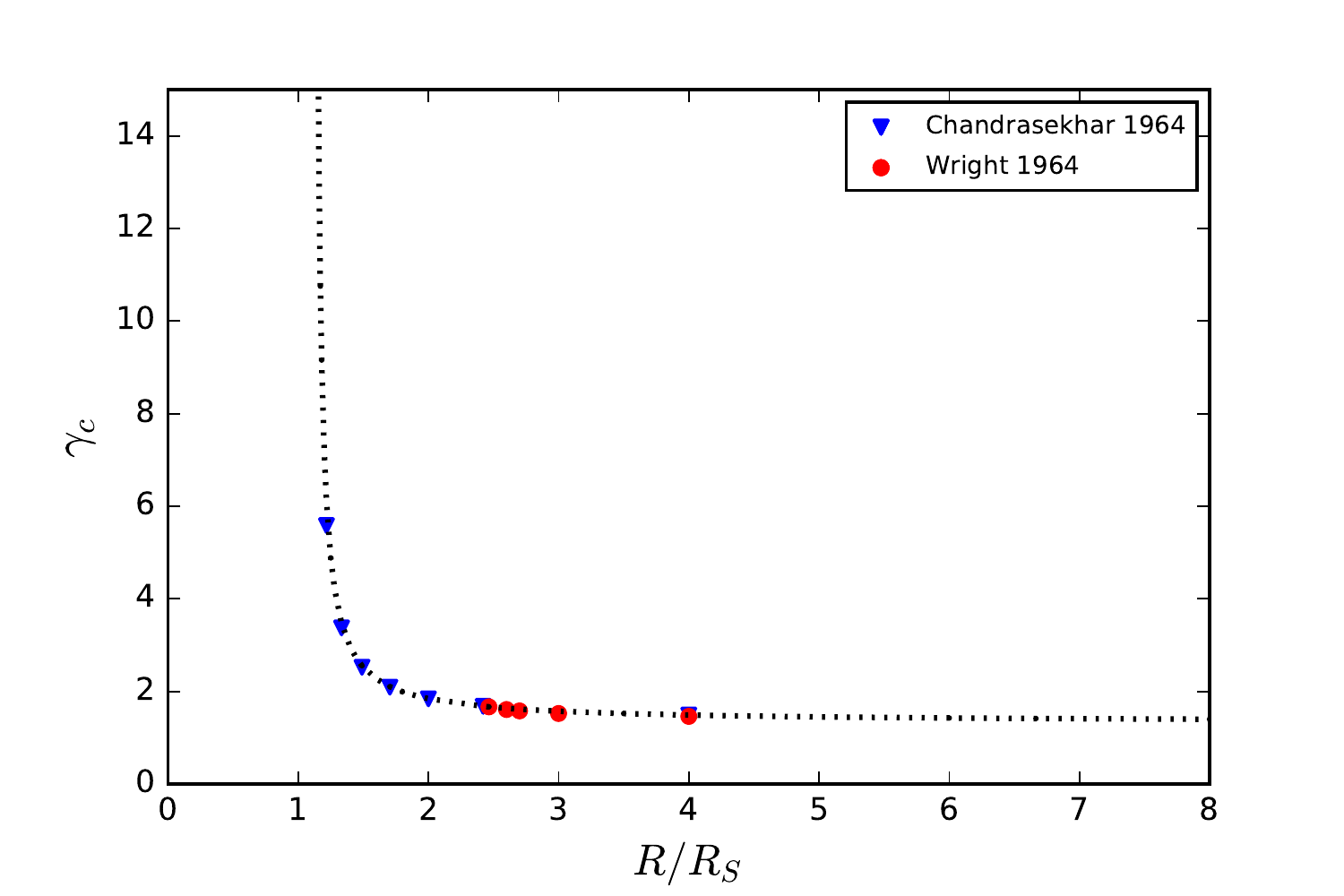}
\caption{\label{fig3} Values of $\gamma_{\rm c}$ (dotted line) as a function of the parameter $R/R_{\rm S}$ for a Schwarzschild star in the regime $R/R_{\rm S}>9/8$. In the limit when $R/R_{\rm S}\to\infty$, $\gamma_{\rm c}$ approaches the Newtonian value 4/3; the star becomes unstable as $R/R_{\rm S}\to9/8$. Our results are in good agreement with those found by Chandrasekhar~\cite{chandra1964b} and Wright~\cite{wright1964}.}
\end{figure}

\begin{figure}
\centering
\includegraphics[scale=0.6]{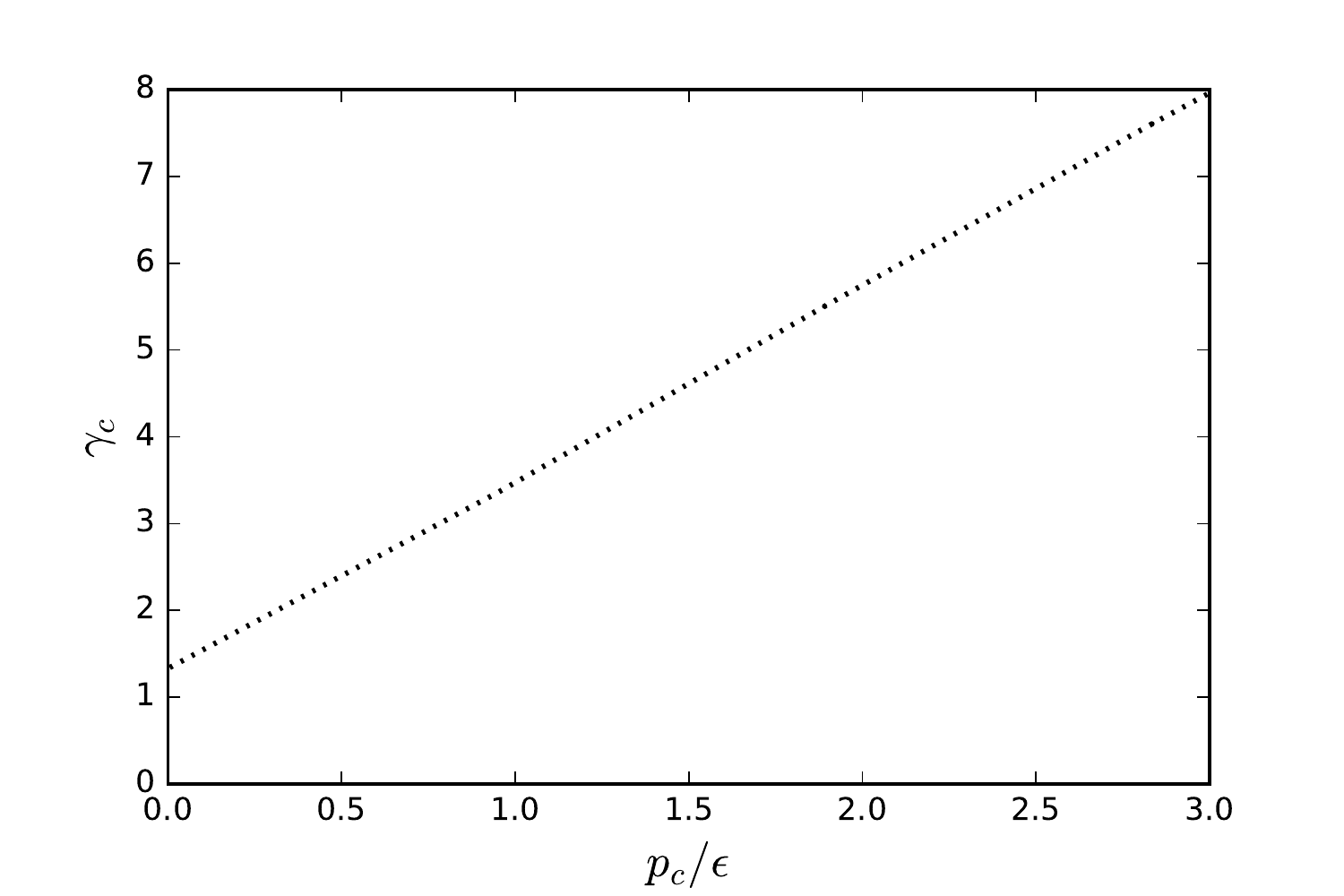}
\caption{\label{fig4} Values of $\gamma_{\rm c}$ as a function of the central pressure $p_{\rm c}$ (in units of $\epsilon$), for various values of the parameter $R/R_{S}$ above the Buchdahl bound. Here the Newtonian value 4/3 is obtained as $p_{\rm c} \to 0$.}
\end{figure}

In \fref{fig4} we plot $\gamma_{\rm c}$ as a function of the central pressure, which is being measured in units of the constant mass-energy density $\epsilon$. In this representation we find the Newtonian limit $\gamma_{\rm c} = 4/3$ as $p_{c}\to 0$, and we can see that $\gamma_{\rm c}$ grows linearly with the central pressure of the star. These results are once again in good agreement with those in \cite{moustakidis2017}.

\begin{figure}
\centering
\includegraphics[scale=0.6]{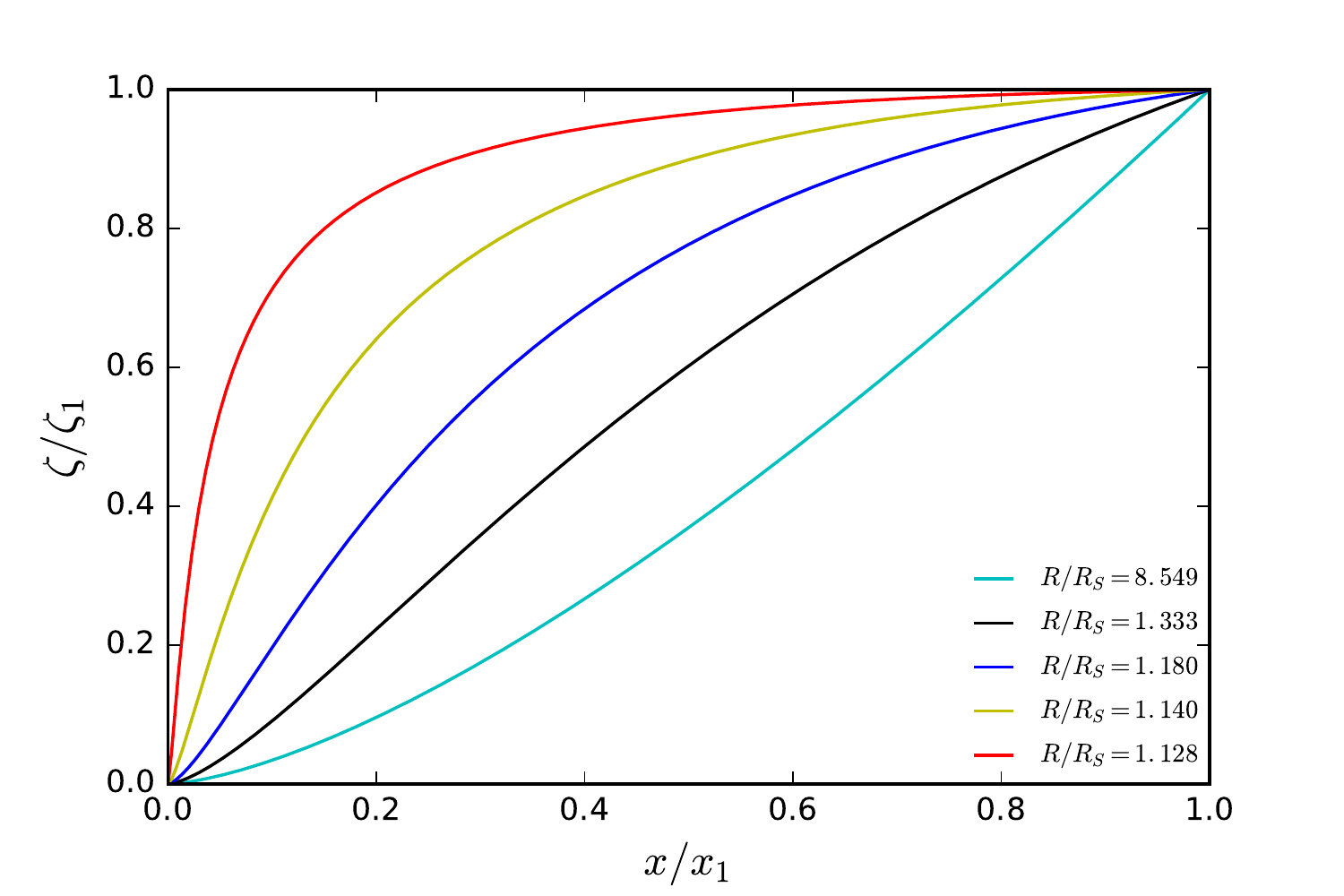}
\caption{\label{fig5} Normalized eigenfunctions $\zeta/\zeta_{1}$ corresponding to the fundamental mode $\omega_{0}^2$, for different values of the parameter $R/R_{\rm S}>9/8$, as a function of $x/x_{1}$. The oscillations become stronger as the stars approach the Buchdahl limit $R/R_{\rm S} \to 9/8$.}
\end{figure}

In \fref{fig5} we plot the eigenfunctions $\zeta$ (normalised by $\zeta_1 \equiv \zeta(x_1)$) corresponding to the marginally stable mode, as a function of $x/x_{1}$. Note that the pulsations grow faster when the radius of the star approaches the unstable configuration in the Buchdahl limit.
 
\subsection{Radial stability in the ultra compact limit $R_{S}<R<(9/8)R_{S}$}
\label{sect5.3}

After having successfully tested our code with the results presented in the previous section, now we proceed to the analysis of the radial stability of an ultra compact Schwarzschild star in the negative pressure regime, $1<R/R_{\rm S}<9/8$. We computed critical values of $\gamma$ for radial stability and obtained the results presented in \fref{fig6} (see also \tref{table2} in the Appendix).

In \fref{fig6} we plot again $\gamma_{\rm c}$ as a function of the parameter $R/R_{\rm S}$, presenting the results obtained in the ultra compact limit $1<R/R_{S}<9/8$, together with the results above the Buchdahl limit previously shown in \fref{fig3}. We can see that, quite surprisingly, the ultra compact Schwarzschild stars beyond the Buchdahl limit become stable again as configurations move away from the instability at $R/R_{S} = 9/8$. It is also interesting to note that the stars are stable even as they approach the Schwarzschild black hole compactness and, as can be seen in the inset of \fref{fig6}, we found that $\gamma_{\rm c} \to 2$ as $R/R_{\rm S}\to 1$. {\footnote{It is important to note here that a finite value of $\gamma_{c}$ on its own does not imply stability: the star must have $\gamma > \gamma_{c}$ for it to be stable. This is merely \emph{allowed} by a finite value of $\gamma_{c}$. However, it should be easily realised in the incompressible fluid of a constant density star - but the situation could be less certain in a more realistic model.}}

\begin{figure}
\centering
\includegraphics[scale=0.6]{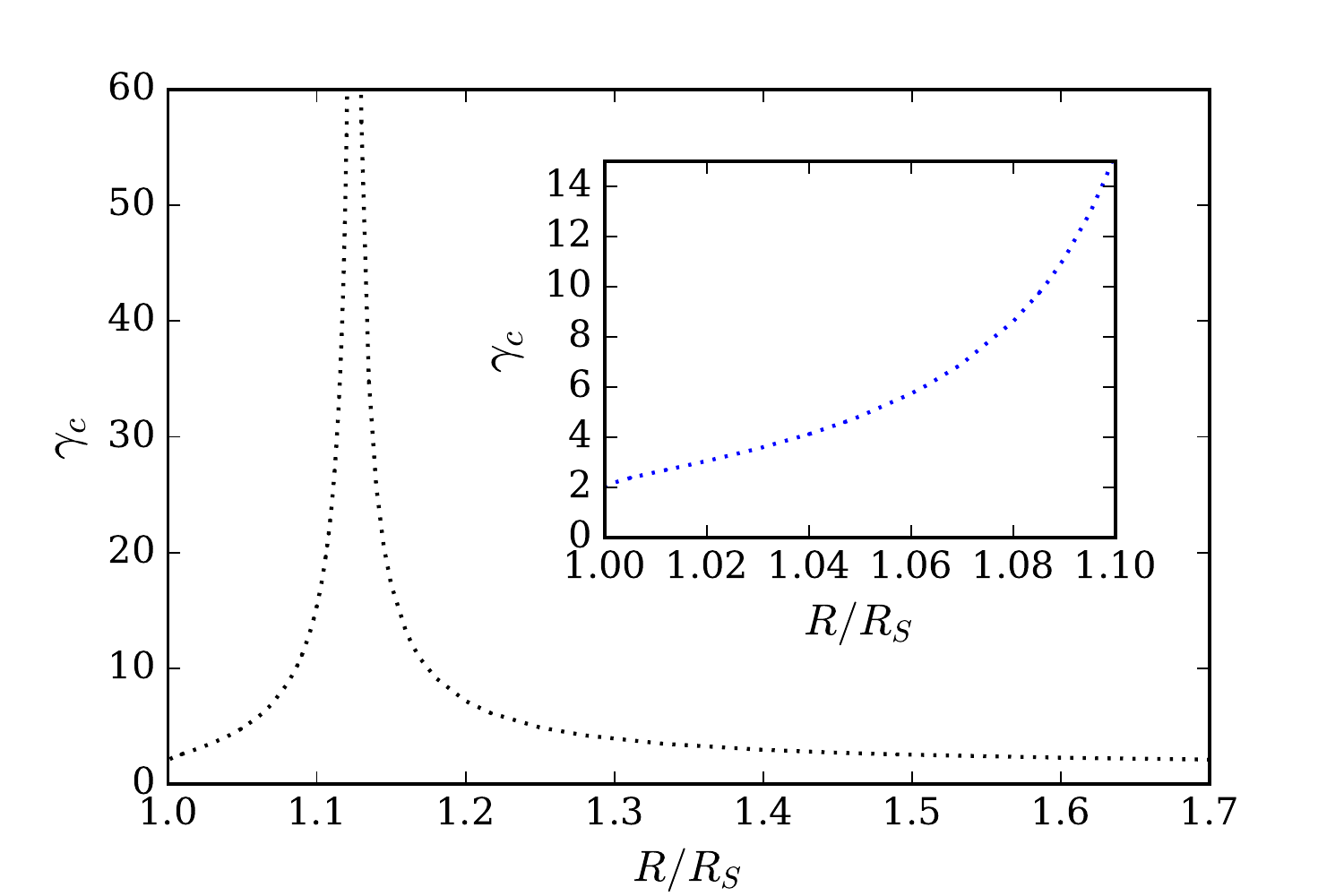}
\caption{\label{fig6} Same as \fref{fig3}, for Schwarzschild stars above and below the Buchdahl bound. After the classical instability at $R/R_{\rm S} = 9/8$, the stars become stable again. The inset shows the behavior of $\gamma_{\rm c}$ in the gravastar limit $R/R_{\rm S} \to 1$.}
\end{figure}

\begin{figure}
\centering
\includegraphics[scale=0.6]{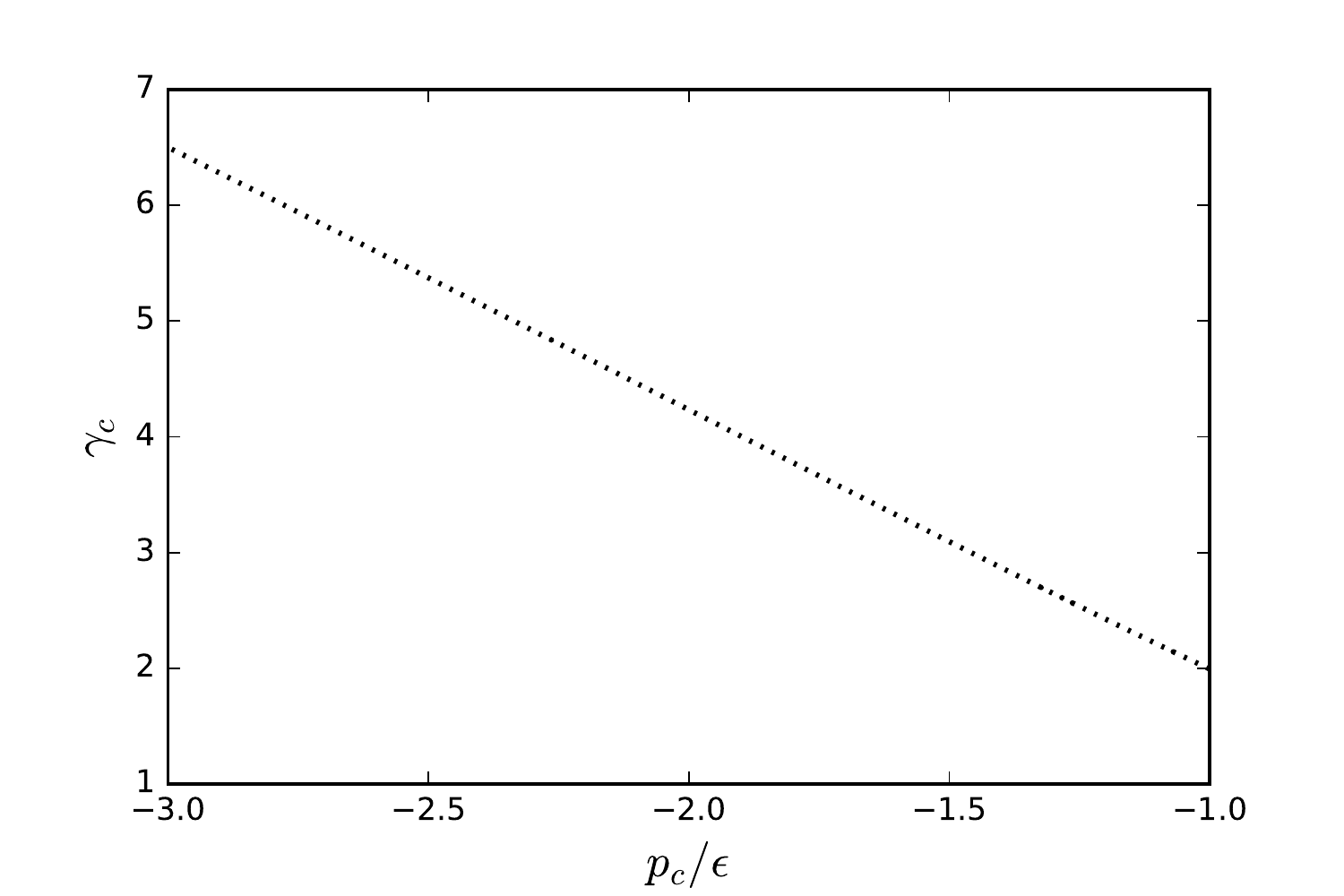}
\caption{\label{fig7} Critical values of $\gamma$ as a function of the central pressure $p_{\rm c}$ (in units of the energy density $\epsilon$) for a Schwarzschild star in the negative pressure regime $R_{\rm S}<R<(9/8)R_{\rm S}$. Note that $\gamma_{\rm c} \to 2$ as $p_c \to -\epsilon$ (that is, as $R\to R_{\rm S}$).}
\end{figure}

This behavior can also be seen in \fref{fig7}, where we plot the critical adiabatic index $\gamma_{\rm c}$, as a function of the central pressure (in units of $\epsilon$), for an ultra compact Schwarzschild star in the negative pressure regime $1<R/R_{\rm S}<9/8$. Note that $p_c \to -\epsilon$ as $R\to R_{\rm S}$ and, in this limit, the interior of the star is entirely described by $p = -\epsilon$ \cite{mazur2015}.

\begin{figure}
\centering
\includegraphics[scale=0.6]{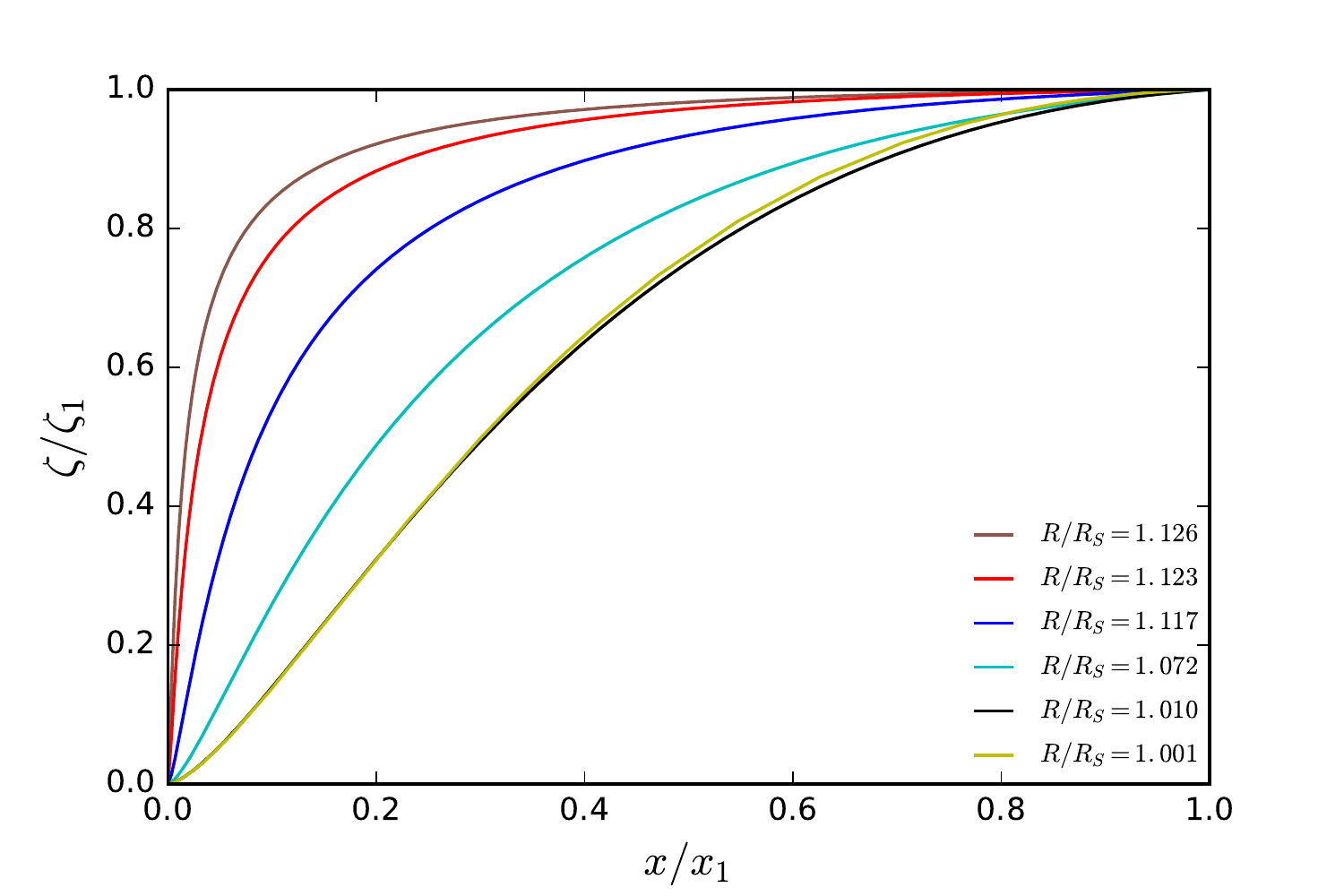}
\caption{\label{fig8} Normalized eigenfunctions $\zeta/\zeta_{1}$ corresponding to the fundamental mode $\omega_{0}^2$, for different values of the parameter $R/R_{\rm S}<9/8$, as a function of $x/x_{1}$. We have included the eigenfunction for the value $R/R_{\rm S}=1.126$, which is above Buchdahl, to show the trend in the attenuation of the eigenfunctions as $R/R_{\rm S} \to 1$.}
\end{figure}

In \fref{fig8} we plot the normalized eigenfunctions $\zeta/\zeta_{1}$, as a function of $x$ (in units of $x_{1}$, its value at the boundary), for an ultra compact Schwarzschild star for various values of the parameter $R/R_{\rm S}<9/8$. Note how the eigenfunctions become more attenuated when the star moves away from the unstable configuration $R/R_{\rm S} = 9/8$ and approaches the limit $R/R_{\rm S}\to 1$. 

\section{Conclusions}
\label{sect6}

In this paper, we addressed the issue of the radial stability of an ultra compact Schwarzschild star, in the recently discovered and still largely unstudied negative pressure regime for $R_{S}<R<(9/8)R_{\rm S}$. This work extends the results found in \cite{chandra1964a} and complements the analysis of radial stability of continuous pressure gravastars presented in \cite{horvat2011}.

Following the theory of infinitesimal, and adiabatic, radial oscillations developed by Chandrasekhar, we determined the critical values of the adiabatic index $\gamma_{\rm c}$ for the onset of instability, above and below the Buchdahl limit. The main result of our investigation is that, in contrast to the usual expectation found in previous works, stability is regained as $\gamma_{\rm c}$ decreases monotonically in the regime $R_{\rm S}<R<(9/8)R_{\rm S}$ and approaches a finite value as $R/R_{\rm S} \to 1$, or gravastar limit.

The dynamical process through which such an ultra-compact star in the negative pressure regime could be formed is still unclear, and remains the biggest challenge for the physical viability of this and related models. Future developments in quantum gravity may provide clues to this still unknown mechanism, and future gravitational wave detections may offer evidence to their existence as black hole alternatives.

\ack

We are thankful to the anonymous referees for their valuable comments and suggestions to improve this manuscript. We are also grateful to Luciano Rezzolla, Pawel Mazur and Emil Mottola for reading an earlier version of the manuscript and providing important comments and suggestions. We also thank Cole Miller and John C. Miller (unrelated) for useful discussions. This work was started when one of us (C.P.) was a visiting researcher at the Center for Mathematics, Computation and Cognition at UFABC in Brazil. This work was supported in part by the S\~{a}o Paulo Research Foundation (FAPESP grant 2015/50421-8) and by the Max Planck Society. C.P. gratefully acknowledges also the support of the Erasmus Mundus Program and the Institute of Physics and Research Centre of Theoretical Physics and Astrophysics, at the Silesian University in Opava, where this work was completed. 

\appendix
\section*{Appendix}
\label{appendix}
\setcounter{section}{1}

In this section we list some of the critical values of $\gamma$ for the radial stability of a constant energy density star, or Schwarzschild star, in the classical regime for lower compactness (\tref{table1}) and in the new negative pressure regime beyond the Buchdahl limit (\tref{table2}). 

\begin{table}
\begin{indented}
\item[]\begin{tabular}{@{} l | l | l }
\br
\hline\hline 
$R/R_{\rm S}$ & $\gamma_{\rm c}$ & $\gamma_{\rm Ch}$\\
\mr
8.549 & 1.394010 & 1.3940 \\
4.000 & 1.489546 & 1.4890 \\
2.420 & 1.678665 & 1.6769 \\
2.000 & 1.843456 & 1.8375 \\
1.704 & 2.103268 & 2.0899 \\
1.490 & 2.554324 & 2.5204 \\
1.333 & 3.475805 & 3.3703\\
1.217 & 6.125634 & 5.5802\\
\br
\end{tabular}
\end{indented}
\caption{\label{table1} Critical adiabatic index $\gamma_{\rm c}$ for radial stability of a Schwarzschild star for various values of the parameter $R/R_{\rm S}>9/8$. Our values $\gamma_{\rm c}$ are in very good agreement with the values $\gamma_{\rm Ch}$ found by Chandrasekhar (see Table 1 in \cite{chandra1964b}). The relative difference increases for decreasing $R/R_{\rm S}$ up to approximately $9\%$. It is important to remark that Chandrasekhar evaluated $\gamma_{\rm Ch}$ via two different trial functions, which provided different results when $R/R_{\rm S}\to 9/8$. His results are expected to be more accurate for less compact stars $R/R_{\rm S} \gg 1$, as in this limit the trial functions approach the true  eigenfunctions.}
\end{table}

\begin{table}
\begin{indented}
\item[]\begin{tabular}{@{} l | l }
\br
\hline\hline 
$R/R_{\rm S}$ & $\gamma_{\rm c}$\\  
\mr
1.120  & 56.211645 \\
1.114  & 31.020834 \\
1.110  & 24.059359 \\
1.085  & 9.742271 \\
1.050  & 4.836920 \\
1.010  & 2.609570 \\
1.001  & 2.142780 \\
1.0001 & 2.037611 \\
\br
\end{tabular}
\end{indented}
\caption{\label{table2} Critical adiabatic index $\gamma_{\rm c}$ for radial stability of an ultra compact Schwarzschild star for various values of $R/R_{\rm S}$, in the regime $1<R/R_{\rm S}<9/8$.}
\end{table}
\newpage
\section*{References}
\bibliography{main}
\bibliographystyle{iopart-num.bst}

\end{document}